\documentclass[final,3p,times,2021]{elsarticle}

\journal{Nuclear Instruments and Methods in Physics Research A}

\usepackage{amssymb}
\usepackage{xspace}
\usepackage[colorinlistoftodos]{todonotes}
\usepackage{multicol}
\usepackage{xcolor}
\usepackage{graphicx}
\graphicspath{ {./figures/} }
\usepackage{caption}
\usepackage{subcaption}
\usepackage{lipsum}

  \usepackage{lineno}

\usepackage[figuresright]{rotating}

\begin{document}

\begin{frontmatter}

\title{Reduction of ion back flow using a quadruple GEM detector with various gas mixtures}

\author[]{Sourav Tarafdar\corref{cor1}}
\ead{sourav.tarafdar@vanderbilt.edu}
\author[]{Senta V. Greene\corref{cor1}}
\ead{senta.v.greene@vanderbilt.edu}
\author[]{Julia Velkovska\corref{cor1}}
\ead{julia.velkovska@vanderbilt.edu}
\author[]{Brandon Blankenship}
\author[]{Michael Z. Reynolds}

\cortext[cor1]{Corresponding authors}
\address{Department of Physics and Astronomy,Vanderbilt University, PMB 401807, 2301 Vanderbilt Place, Nashville, TN 37235-1807, U.S.A}

\begin{abstract}
In gaseous tracking detectors with a large gaseous volume multiple layers of Gas Electron Multipliers (GEM) can be used to block positive ions from flowing back into the active volume, which is detrimental to the tracking performance. We report on studies of effective gain, ion backflow (IBF), and energy resolution in quadruple GEM detectors, and on strategies for minimizing IBF by optimizing the operating voltages of the individual GEM layers and the potential differences between different layers.
\end{abstract}


\end{frontmatter}

\section{Introduction}
\label{sec:intro}
Gas Electron Multipliers (GEMs)~\cite{SAULI1997531} are widely used in high
energy particle and nuclear experiments to track charged particles. GEMs are
radiation hard, can amplify small signals, and have excellent energy
resolution~\cite{SAULI20162}.

Key parameters that
determine the quality of GEM detectors are effective gain, energy resolution, spatial resolution, time resolution, and ion backflow (IBF). The effective gain is defined to be the ratio of the total number of avalanche electrons after all gain elements (GEMs in this article) and the total number of primary electrons created through ionization in the drift volume. IBF is the fraction of the  positive ions that have traveled back to the drift cathode. Some of the factors that affect these parameters are the operating voltages for the GEMs, the gas medium in which the detector is operated, 
and the pitch of the holes in the GEM layers~\cite{SAULI1997531,SAULI20162, BRESSAN1999254,ZIEGLER2001260,ZHANG2018184, Greene:2020iei}. This article provides measurements of the effective gain, IBF, and energy resolution with different gas mixtures and different voltage configurations for a quadruple GEM detector. Gaseous tracking detectors in a high multiplicity environment, such as a Time Projection Chamber (TPC) ~\cite{ALME2010316, ANDERSON2003659}, typically have a large drift volume. The drift region gets populated by positive ions because of their low drift velocity. The accumulation of these positively-charged ions, known as space charge, distorts the desired uniform electric field, causing inaccurate charged-particle tracking. In TPCs with GEM readout~\cite{ALME2010316, Dehmelt:2018yev, Tarafdar:2018bku} IBF can contribute significantly to the space charge, so it is important to suppress IBF without compromising other important operation parameters, such as gain and resolution. We report on the measurements of the effective gain, energy resolution, and IBF of a quadruple GEM detector in several different gas mixtures, and discuss a configuration of operating voltages of the individual GEM layers and the potential differences between different layers that minimizes IBF, while keeping the effective gain constant. 

\section{Experimental setup}

\begin{figure}
     \centering
     \begin{subfigure}[!htb]{0.5\textwidth}
         \centering
         \includegraphics[width=\textwidth]{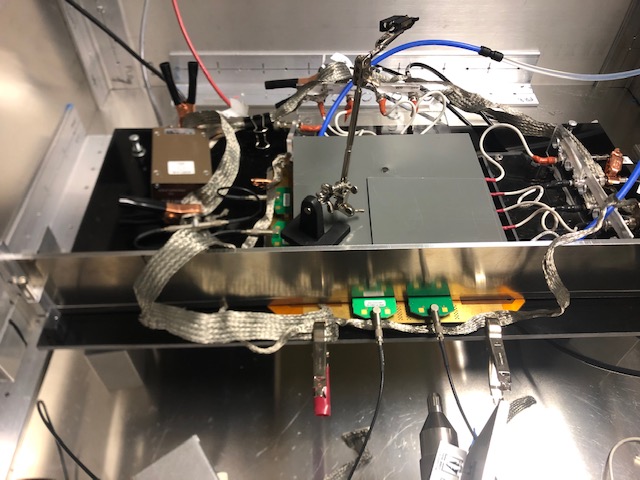}
         \caption{}
         \label{fig:picture}
     \end{subfigure}
    \hfill
     \begin{subfigure}[!htb]{0.85\textwidth}
         \centering
         \includegraphics[width=\textwidth]{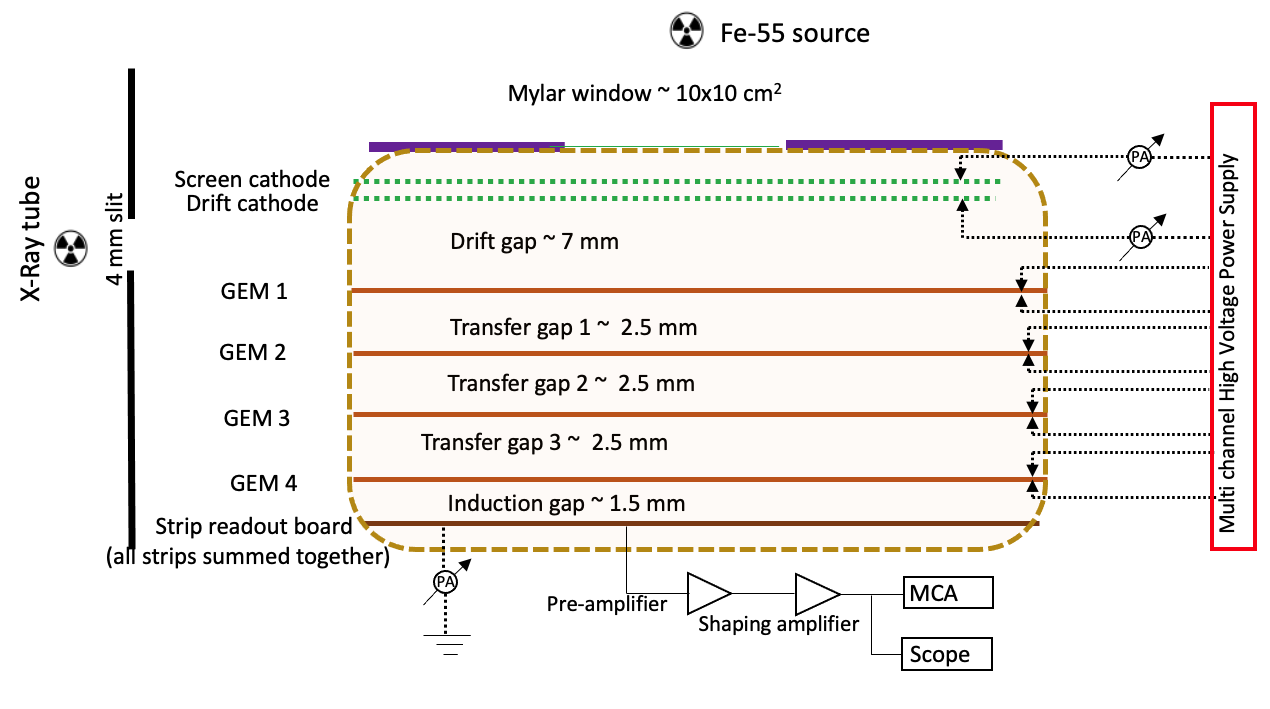}
         \caption{}
         \label{fig:quad_gem}
     \end{subfigure}
        \caption{Experimental setup: a) A photograph of the GEM detector on the experimental test bench.  b) Schematic diagram of the quadruple GEM detector setup.}
        \label{fig:setup}
\end{figure}

The GEM foils used in this study are 
standard GEMs with 140 $\mu$m pitch and
70 $\mu$m  hole diameter with ${\rm 10\times10  cm^{2}}$ active area. Figure~\ref{fig:picture} shows the experimental
setup on the laboratory test bench and Fig.~\ref{fig:quad_gem} is the block diagram of the overall setup of used for the measurements of the effective gain, energy resolution, and IBF of the quadruple GEM detector.

The quadruple GEM detector is a stack of four layers of GEMs spaced 2.5 mm apart and sandwiched between  the drift cathode and the X-Y strip readout board. For the measurements presented in this manuscript, the signal was recorded after summing the 512 X-Y strips of the readout board. The gap between the drift cathode and the top of the first GEM (drift gap) is 7 mm , and the gap between the bottom of the last GEM and the readout board (induction gap) is 1.5 mm. The measurements were performed in ArCO$_{2}$ mixed in different proportions, NeCO$_{2}$(90:10) and NeCF$_{4}$(90:10) gas mixtures. The gases were mixed using an in-house gas mixing system, capable of mixing two different gases with 0.15 \% accuracy in their ratio. The gases used in the experiment were of ultrahigh purity grade (99.999$\%$ pure).  

The quadruple GEM detector's individual electrodes were biased by individual channels of a high voltage power supply via 10 MOhm protection
resistors.  X-rays from a collimated Fe$^{55}$ radioactive source placed perpendicular to the surface of the detector were used for the measurement of the effective gain of the detector. For measuring IBF, a mini
X-ray tube was used with 2 mm collimation. The X-ray tube was placed on the side of the detector and aimed so as to irradiate only the drift gap. A 4 mm wide slit was placed near the sidewall of the detector and along the X-ray path to ensure that the primary ionization took place strictly inside the drift gap. A screen cathode located 0.5 mm above the drift cathode was added to measure the current from the ionization of the gas above the drift gap, in case imperfect alignment between the slit and the drift gap caused X-ray photons to enter the region above the drift gap. The measured current from the screen cathode, if any, would serve as a correction factor to the current measured from ions back-drifting to the drift cathode. 

The readout electronics for the detector signal comprised an Ortec 142B charge sensitive pre-amplifier, an Ortec 760 shaping amplifier, and an Amptek MCA-8000D multi-channel analyzer. A Tetronix MDO3054 oscilloscope was used for initial set-up and additional monitoring. The IBF was estimated based on the measurements of the current in the readout board that is induced by avalanche electrons , and the current in the drift cathode from back-drifting positive ions. The measurements were performed using a highly sensitive Picologic picoammeter.

Throughout these studies, the relative humidity, temperature, gas flow and pressure inside the detector were kept constant. The flow rate for the quadruple GEM detector was set at 100 sccm, while the atmospheric temperature and relative humidity were kept constant at 24 degrees Celsius and 28$\%$, respectively, using digital humidity and temperature controllers. 

\section{Results}
The following subsections shows results on effective gain, energy resolution , and IBF. The suppression of  IBF in a quadruple GEM detector for various gas mixtures is studied. 
\label{sec:results}
\subsection{Effective gain and energy resolution}

\label{subsec:effgain}
 
The effective gain, $G_{\mathrm{eff}}$, is given by the ratio of the number of avalanche electrons, N$_\mathrm{av}$ collected at the readout board and the number of primary electrons N$_\mathrm{p}$ produced inside the drift gap region from the passage of primary ionizing particles. The term effective gain instead of absolute gain has been used here because avalanche electrons that become attached to the dielectric layer of each GEM do not contribute to the collected signal. 

The gain measurement should be decoupled from any amplification in the readout electronics, and therefore we first calibrated the electronics response to a known amount of input charge. This was done by replacing the quadruple GEM detector in the actual setup by a precise $C=1$~pF capacitor with 2$\%$ tolerance.
\begin{figure}
     \centering     
     \begin{subfigure}[htb!]{0.8\textwidth}
         \centering
       \includegraphics[width=\textwidth]{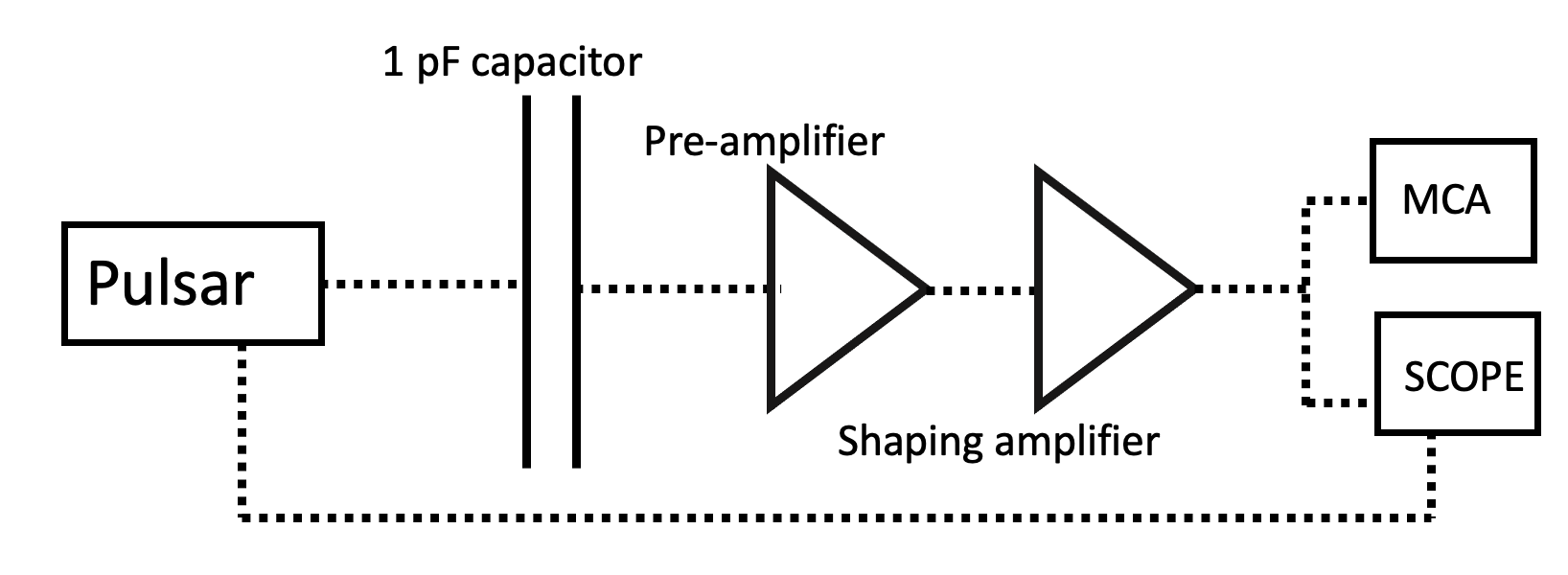}
         \caption{}
         \label{fig:cal_scheme}
     \end{subfigure}
     \hfill
     \begin{subfigure}[!htb]{0.8\textwidth}
         \centering
         \includegraphics[width=\textwidth]{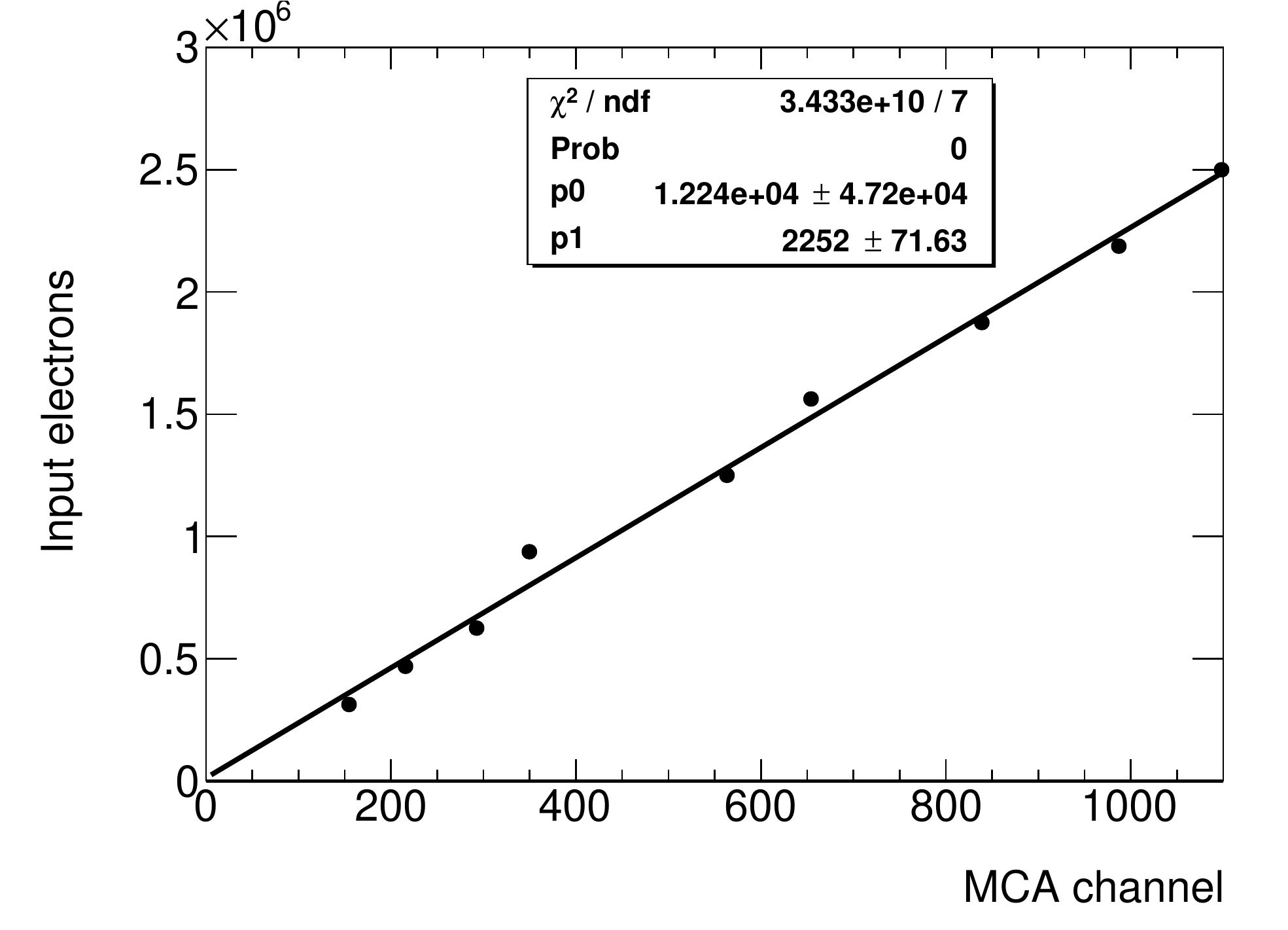}
         \caption{}
         \label{fig:cal_curve}
     \end{subfigure}
\caption{ a) Schematic diagram of the electronics calibration, and  b) Calibration curve }
        \label{fig:cal_est_setup}
\end{figure}

The schematic diagram of the measurement set-up is shown in Fig.~\ref{fig:cal_scheme} (a). Using the pulser, a known charge Q$_{\mathrm{input}} = C \times V_{\mathrm{input}}$ was injected into the readout electronics and the output pulse was registered in the MCA. The correlation between input charge and the corresponding channel number of the MCA was obtained. The resulting calibration curve is shown in Fig.~\ref{fig:cal_curve} (b). 

The 5.9 keV X-rays from a Fe$^{55}$ radioactive source were used to measure $G_{\mathrm{eff}}$. For a given gas mixture, the number of primary electrons generated by the absorption of these X-rays in the drift region of the detector was calculated using a simple composition law and the mean ionization energy for each gas component, $W_{i}$.
\begin{equation}
    \label{eqn:prim_ion_e}
   \mathrm{N_{{p}} = \frac{5.9 keV}{f_{A}\times W_{i}^{A} + f_{B}\times W_{i}^{B}}}, 
\end{equation}
where $f_{A}$ and $f_{B}$ are the respective fractions of gas A and gas B in the mixture. The values of $W_{i}$ used in the calculations are given in Table~\ref{table:1}. Table~~\ref{table:2} gives the values of N$_\mathrm{p}$ for the gas mixtures used in the experiment. 
 \begin{table}[h!]
\centering
 \begin{tabular}{||c | c| c| c| c ||} 
 \hline
 Gas  & Ar & Ne & CO$_{2}$ & CF4 \\ [0.5ex] 
 \hline\hline
 W$_{i}$ (eV) &  26 & 36 & 33 & 54\\[1ex]
 \hline
 \end{tabular}
 \caption{Mean ionization energy W$_{i}$ for different gases used in the experiment. The values for Ar, Ne and CO$_{2}$ are taken from ~\cite{Sauli:117989}, whereas for CF4 the value is from ~\cite{SHARMA}. }
\label{table:1}
\end{table}
\begin{table}[h!]
\centering
 \begin{tabular}{||c | c ||} 
 \hline
 Gas mixture & Number of primaries \\ [0.5ex] 
 \hline\hline
 ArCO$_{2}$-70:30 & 210\\ 
 \hline
 ArCO$_{2}$-80:20 & 215 \\
 \hline
 ArCO$_{2}$-90:10 & 221 \\ 
 \hline
 NeCO$_{2}$-90:10 & 164 \\ 
 \hline
 NeCF4-90:10 & 155 \\[1ex]
 \hline
 \end{tabular}
 \caption{Number of primary electrons in different gas mixtures produced by   5.9 keV X-rays from a Fe$^{55}$ source that have been completely absorbed in the drift region of the quadruple GEM detector.}
\label{table:2}
\end{table}

\begin{figure}[!htb]
\begin{center}
\includegraphics[width=0.7\textwidth] {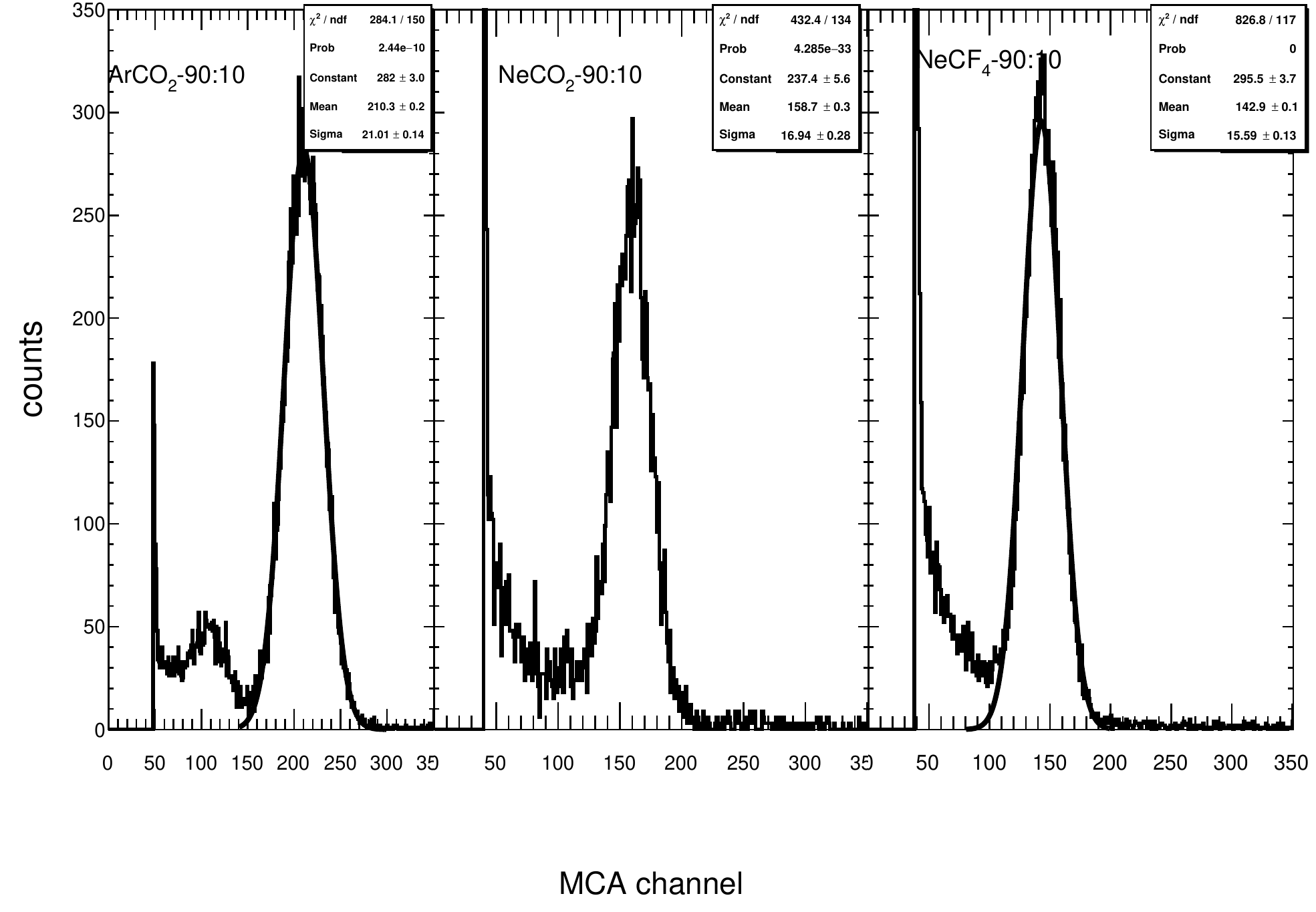}
\caption{Fe$^{55}$ spectra measured by quadruple GEM detector for Ar- and Ne-based gas mixtures at effective gain of 2000.} 
\label{fig:fe55_spectra}
\end{center}
 \end{figure}

To calculate $G_{\mathrm{eff}}$ we need to measure the average number of electrons N$_\mathrm{av}$ collected at the readout board of the quadruple GEM detector corresponding to a 5.9 keV photon. Figure~\ref{fig:fe55_spectra} shows the Fe$^{55}$ spectra measured in the MCA with different operating gas mixtures. The main feature of this spectrum is the 5.9 keV X-ray peak. In the Ar gas mixture, a second peak with energy around 3 GeV is also visible. This so called "escape peak" arises when the 5.9 keV photon is absorbed by an inner-shell electron and the vacancy is filled by an upper-shell electron, which emits a characteristic X-ray (in Ar $K_{\alpha} = 2.9$ keV) that leaves the gas. In Ne, the characteristic X-rays are much softer and are readily absorbed, so the escape peak is not visible in the measured spectra. 

 The 5.9 keV X-ray peaks for all gas mixtures for the quadruple GEM were fitted with Gaussian functions. The mean values of the fit were then converted to N$_\mathrm{av}$ using the slope of the calibration curve in Fig.~\ref{fig:cal_curve}. The effective gain of the quadruple GEM detector was then calculated using the measured N$_\mathrm{av}$ and the calculated values for number of primary electrons N$_\mathrm{p}$ from Table~\ref{table:2}.
 
\begin{figure}[!htb]
\begin{center}
\includegraphics[width=1.\textwidth] {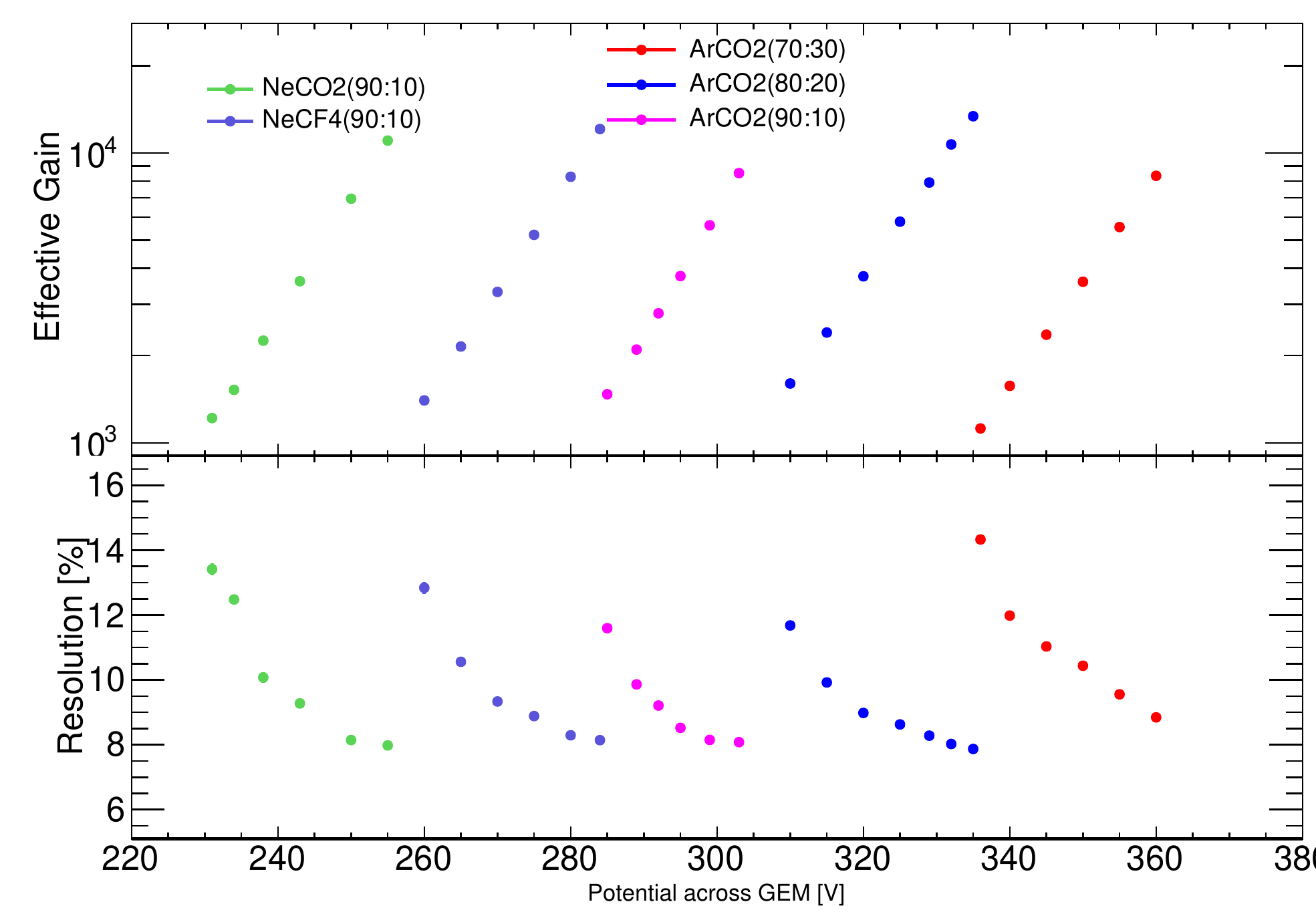}
\caption{Effective gain (top panel) and corresponding energy resolution (bottom panel) for the quadruple GEM detector using Argon- and Neon-based gas mixtures as functions of the potential difference across each GEM.} 
\label{fig:gain_res}
\end{center}
\end{figure}

The top panel of Fig.~\ref{fig:gain_res} shows the results obtained in different gas mixtures while varying the potential difference across each GEM. With the increase in potential difference across each GEM, the electric field in the holes increases and hence the  number of avalanche electrons and effective gain. 

 The bottom panel of Fig.~\ref{fig:gain_res} shows the energy resolution of the detector determined as the ratio of the Gaussian width and mean of the measured 5.9 keV peaks. 

The bottom panel of Fig.~\ref{fig:gain_res} shows the energy resolution for the Quadruple GEM detector for different potential differences across each GEM. 

\subsection{Ion backflow}
The ion backflow (IBF) is the fraction of positive ions traveling back to the drift region of the GEM detector. Experimentally, it is determined as the ratio of the  current measured at the cathode and the anode. The measurements were done using the X-ray tube as the ionization source. The measured cathode and anode currents were corrected for background current and for ionization outside the drift region, as follows:  

\begin{equation}
    \label{IBF_eqn}
   \mathrm{IBF = \frac{I_{\mathrm{cathode}}^{\mathrm{X-ray}}- I_{\mathrm{screen}}^{\mathrm{X-ray}}- I_{\mathrm{cathode}}^{\mathrm{w/o X-ray}}}{I_{\mathrm{anode}}^{\mathrm{X-ray}}- I_{\mathrm{anode}}^{\mathrm{w/o X-ray}}}}
\end{equation}
where  I$_{\mathrm{cathode}}^{\mathrm{X-ray}}$ is the current measured from the drift cathode from back-flowing ions with the X-ray tube on, I$_{\mathrm{screen}}^{\mathrm{X-ray}}$ is the current measured from the screen cathode from primary ionization of gas in the region above the drift cathode, I$_{\mathrm{anode}}^{\mathrm{X-ray}}$ is the current measured from the anode from avalanche electrons with X-ray tube on, and I$_{\mathrm{cathode}}^{\mathrm{w/o X-ray}}$ and I$_{\mathrm{anode}}^{\mathrm{w/o X-ray}}$ are the background currents at the cathode and the anode, respectively.

\begin{figure}[!htb]
\begin{center}
\includegraphics[width=0.70\textwidth] {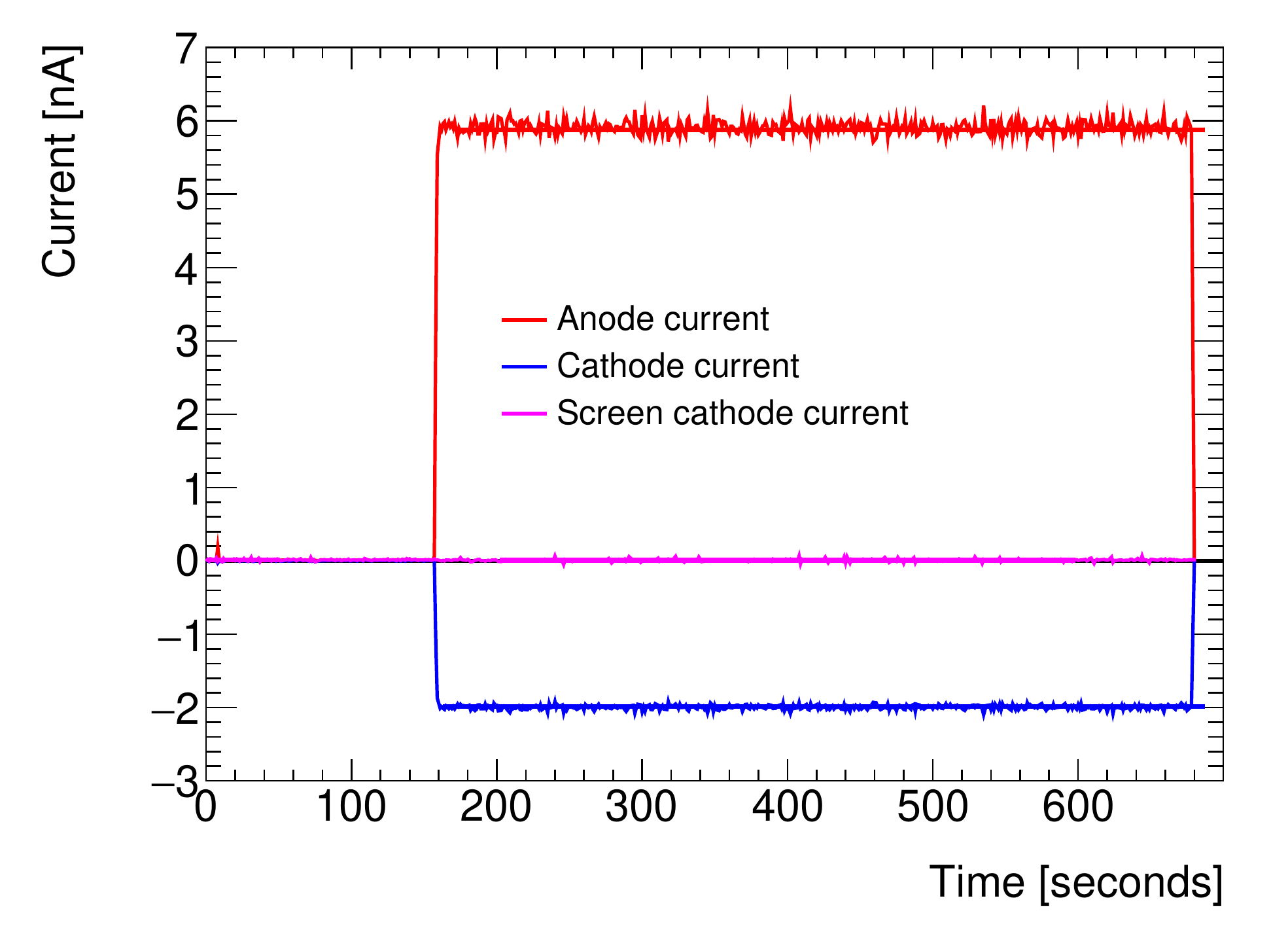}
\caption{Measured current from anode, cathode, and screen cathode for quadruple GEM detector with effective gain of 3500 in ArCO$_{2}$(90:10) gas, both before and after turning on the X-ray tube} 
\label{fig:an_cat_sc_I}
\end{center}
\end{figure}

Figure~\ref{fig:an_cat_sc_I} shows the measured currents from the anode, cathode, and screen cathode for an extended period of time using the quadruple GEM detector filled with the ArCO$_{2}$(90:10) gas mixture and operated with an effective gain of about 3500.  The step in the measured currents indicates the time when the  X-ray tube was turned on. The current measured at the screen cathode was consistent with zero, which indicates that the alignment between the X-ray tube and the quadruple GEM detector was good, and primary ionization happened strictly in the drift region as intended. 

Measurements from the anode, drift cathode, and screen cathode as shown in Fig.~\ref{fig:an_cat_sc_I} were fit with straight lines for the periods when the X-ray tube was on and off to determine the corresponding currents in Eqn.~\ref{IBF_eqn}. The background currents  I$\mathrm{_{{cathode}}^{w/o X-ray}}$ and I$\mathrm{_{{anode}}^{w/o X-ray}}$ were of the order of a few pico-amperes.
The measurements were performed in several different gas mixtures. 
\begin{figure}[!htb]
\begin{center}
\includegraphics[width=1.\textwidth] {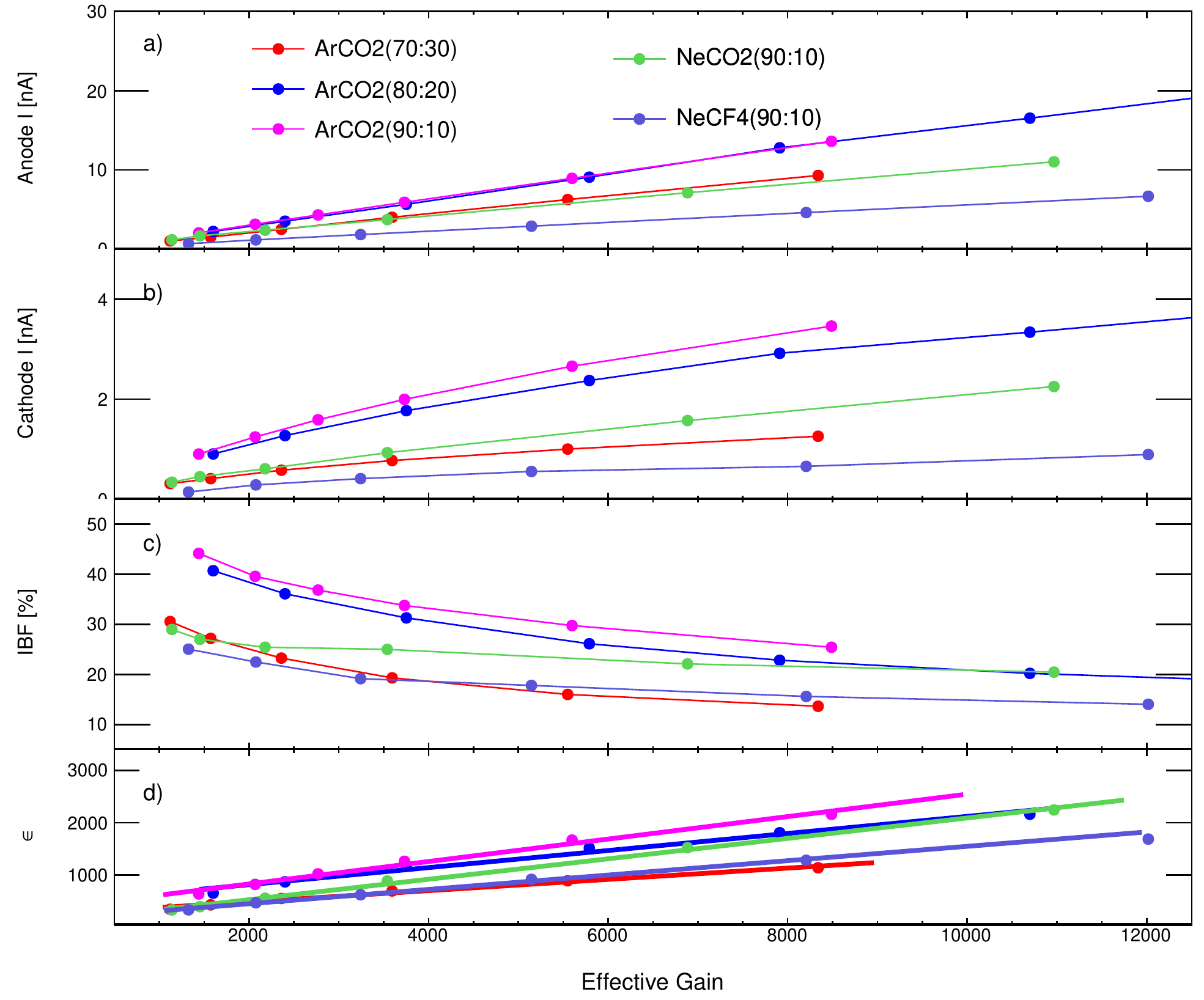}
\caption{From top panel to bottom, as functions of the effective gain: (a) anode current,  (b) cathode current, (c) IBF, (d)  number of back-flowing ions per incoming electron for different gas mixtures used in the quadruple GEM detector. The data points for panels (a),(b), and (c) are connected with line segments, whereas the data points in panel (d) are fit with a line.} 
\label{fig:an_cat_ibf_eps}
\end{center}
\end{figure}
Figure~\ref{fig:an_cat_ibf_eps} shows the current measured from the anode (a), cathode (b), and corresponding IBF (c) for different gas mixtures as functions of the effective gain of the detector. As expected,  with the increase in effective gain the measured current at the drift cathode from back flowing ions increases. However, IBF decreases because the anode current increases faster than the cathode current. If one is concerned about accumulation of space charge and subsequent distortion of the electric field in a TPC, then  the absolute number of back-drifting ions needs to be controlled at a given effective gain. Therefore, it is useful to quantify the number of back-drifting ions per incoming electron, which can be expressed as ${\mathrm{\epsilon = IBF \times Gain_{effective}}}$. This quantity is shown in panel (d) of  Fig.~\ref{fig:an_cat_ibf_eps}. 

For optimal performance, $\epsilon$ and IBF need to be minimized at a fixed value of the effective gain at which the detector is operated.  

\subsection{Ion back flow reduction and its effect on energy resolution}
This section focuses on various ways of reducing IBF and $\epsilon$. The effect of IBF suppression on the energy resolution is also studied for different Ar- and Ne-based gas mixtures. GEM-based detectors in collider experiments are generally operated at a constant effective gain. Currently, it is planned that the GEM-based TPCs for both ALICE and sPHENIX will operate at an effective gain of 2000. This  motivated the studies presented here, in which the potential differences across the GEMs and in the gap fields were chosen to maintain a constant effective gain of 2000 while attempting to keep the  IBF and $\epsilon$ to a minimum. 

A detailed IBF suppression study is done with ArCO$_{2}$(70:30) gas. After optimizing the gap fields and the potential differences across the GEMs for the lowest IBF at an effective gain of 2000, the same operating points were applied to other gas mixtures to investigate the effect on IBF and resolution. 

\begin{figure}[!htb]
\begin{center}
\includegraphics[width=0.7\textwidth] {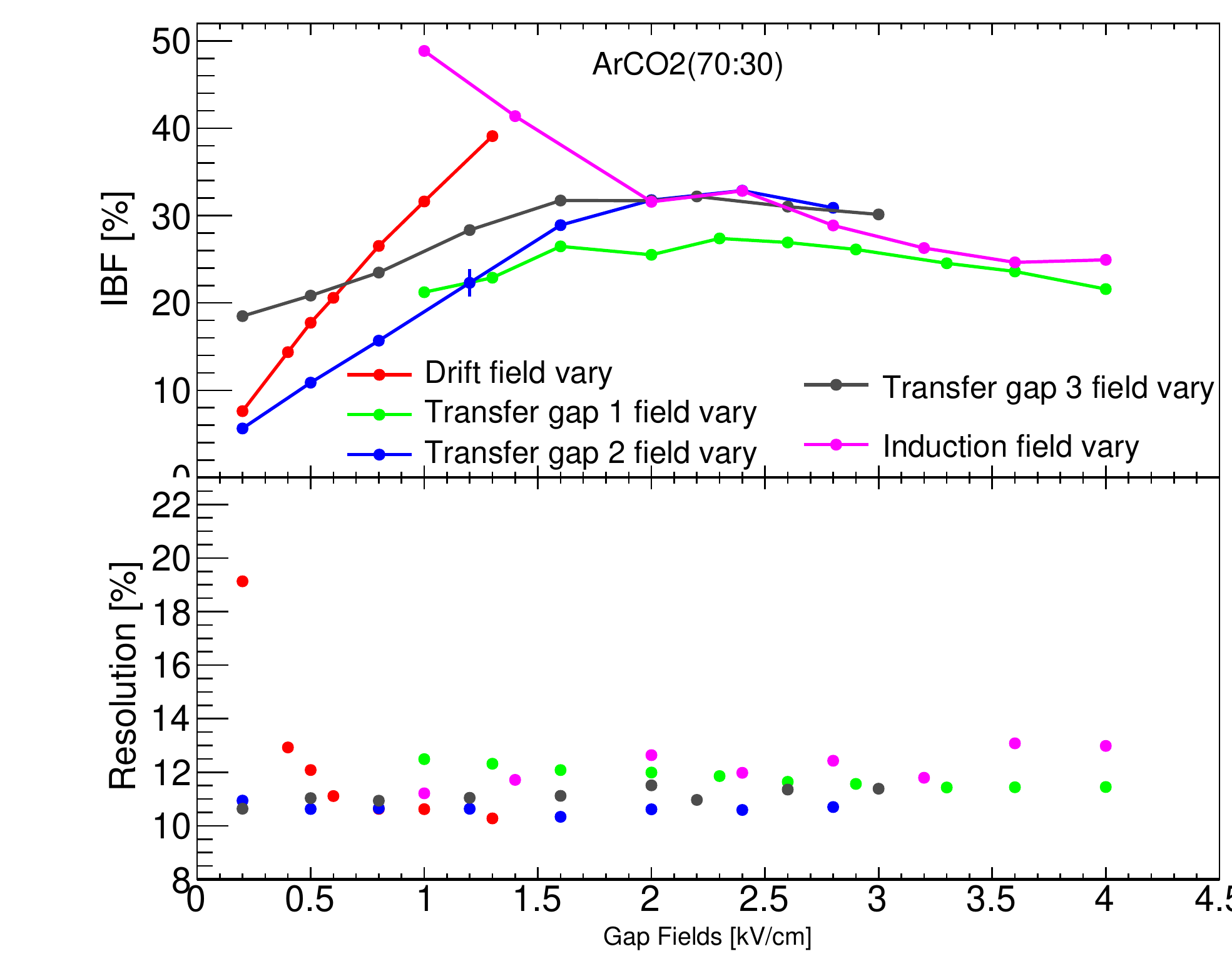}
\caption{IBF (top panel) and energy resolution (bottom panel) versus changes in one specific gap field as denoted in the legend. The potentials across each of the four GEM detectors are the same, with values that are adjusted to maintain an overall effective gain of 2000. } 
\label{fig:ibf_vs_gapfld_dvgemsame}
\end{center}
\end{figure}

As a starting point, all GEM detectors were operated with 345 V potential difference across the GEM so that the effective gain of the quadruple GEM detector was 2000, as shown in the studies in Fig.~\ref {fig:gain_res}. The gap fields were then varied one-by-one, while keeping the other gap fields constant. In order to maintain the same overall effective gain of 2000, the potentials across the GEMs were adjusted accordingly after each gap-field change. All four GEMs were operated at the same gain. The IBF and resolution obtained with the variation of each gap field are shown in Fig.~\ref{fig:ibf_vs_gapfld_dvgemsame}. The lowest IBF values (on the order of 10\%) were achieved when the fields in transfer gap 2 and transfer gap 3 and the drift gap were kept below 1kV/cm, while the fields in transfer gap 1 and the induction gap were higher. The effect on energy resolution is shown in the bottom panel of Fig.~\ref{fig:ibf_vs_gapfld_dvgemsame}. The energy resolution of the detector is affected more by the change in drift gap field as compared to changes in the transfer and induction gap fields. The energy resolution gets worse with a decrease in the drift gap field. This is understandable considering that with a lower electric field in the drift region, the transparency to electrons from primary ionization  is reduced. 

\begin{figure}[!htb]
\begin{center}
\includegraphics[width=0.70\textwidth] {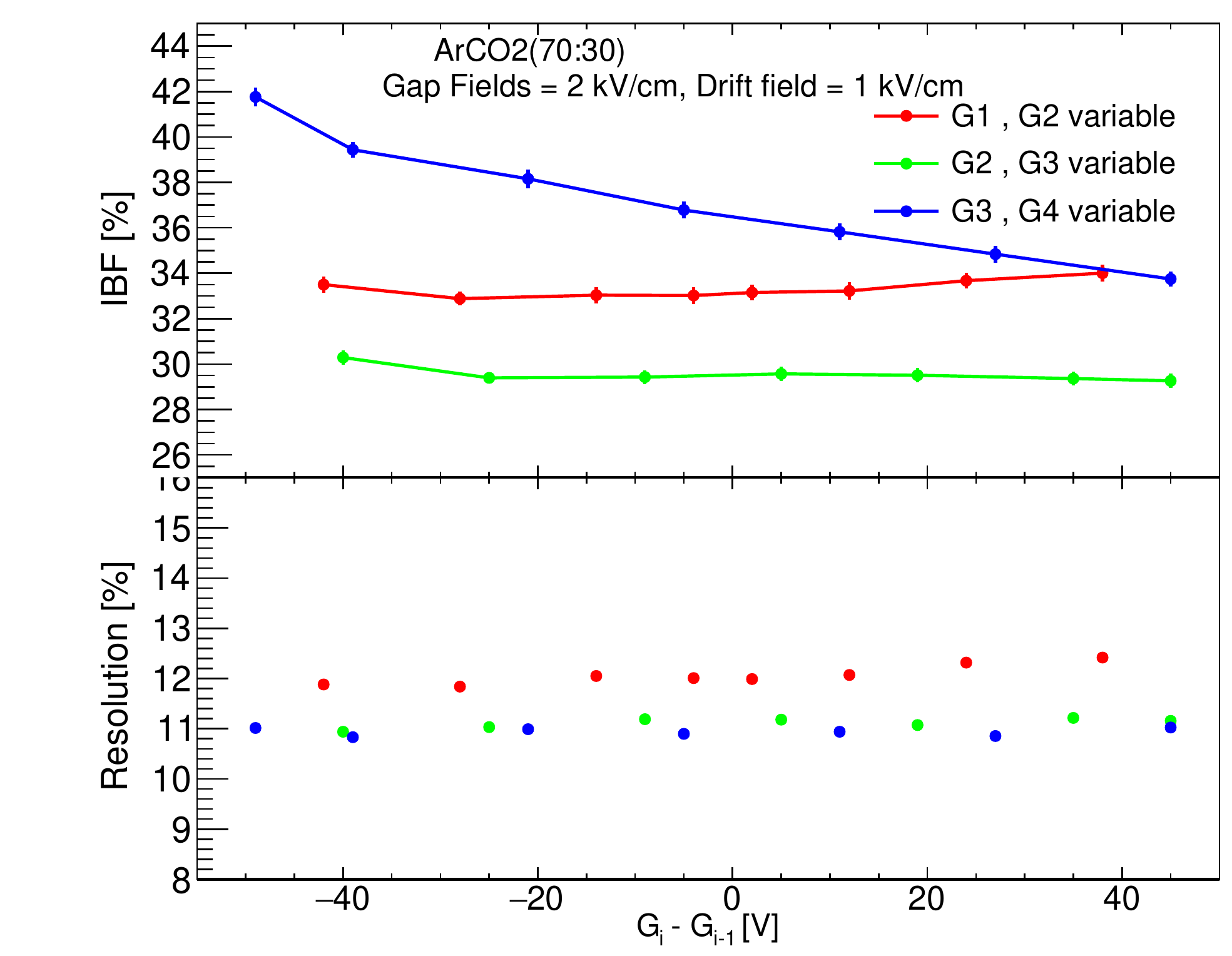}
\caption{IBF and energy resolution of the quadruple GEM detector as a functions of the difference in the potential applied across successive GEMs $G_{i} - G_{i-1}$.  The overall effective gain of the quadruple GEM detector is maintained at about 2000, but the gains of the individual GEMs are not the same. } 
\label{fig:ibf_vs_gapfld_dvgemasymm}
\end{center}
\end{figure}

\begin{table}
\centering
 \begin{tabular}{||c | c ||} 
 \hline
 Gaps & Field [kV/cm] \\ [0.5ex] 
 \hline\hline
 Drift gap & 1.0\\ 
 \hline
 Transfer gap 1 & 3.0 \\
 \hline
 Transfer gap 2  & 0.4 \\ 
 \hline
 Transfer gap 3 & 0.5 \\ 
 \hline
 Induction gap & 4.0 \\[1ex]
 \hline
 \end{tabular}
 \caption{Gap fields in which the lowest IBF was obtained while operating each GEM at a different gain. The effective gain of the quadruple GEM detector was maintained at a value of approximately 2000.}
\label{table:3}
\end{table}
\begin{figure}[!htb]
\begin{center}
\includegraphics[width=0.70\textwidth] {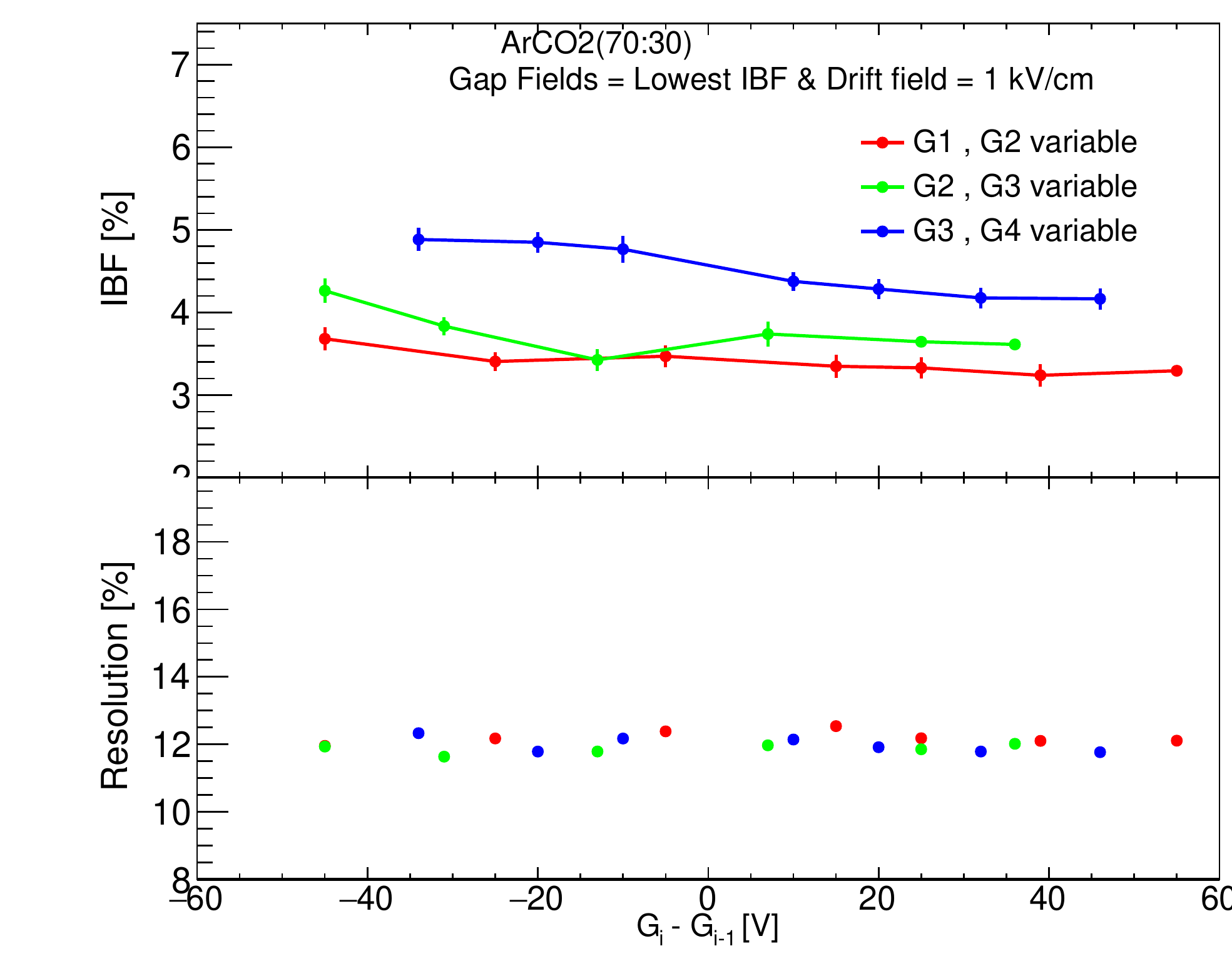}
\caption{IBF and energy resolution of the quadruple GEM detector as a functions of the difference in the potential applied across successive GEMs $G_{i} - G_{i-1}$.  The transfer gap and induction gap fields are set to have minimum IBF. The drift gap field is set at 1 kV/cm. The overall effective gain of the quadruple GEM detector is maintained at about 2000, but the gains of the individual GEMs are different. } 
\label{fig:ibf_vs_gapfld_dvgemasymm_gapfldslowest}
\end{center}
\end{figure}

A study of IBF suppression and energy resolution was also performed by operating the individual GEMs at a different gain while still keeping the effective gain of the detector at about 2000. During this study, the drift gap field was set to 1 kV/cm, while the transfer and induction gap fields were set to 2 kV/cm. Two of the four GEMs were operated at the same gain while the other two GEMs were operated at different gains by changing the potential across each of them so that the total effective gain of the detector remained constant at about 2000. Figure~\ref{fig:ibf_vs_gapfld_dvgemasymm} shows the IBF and energy resolution versus the potential difference applied across two consecutive layers of GEM denoted as $G_{i} - G_{i-1}$, while keeping the gain of the remaining two GEMs unchanged. The IBF remains at a constant value for variations of $G_{2} - G_{1}$ or $G_{3} - G_{2}$. However, the IBF is reduced as $G_{4} - G_{3}$ increases, e.g. if the fourth GEM is operated at a higher gain than the  previous (third) GEM. This is because the field lines in the holes of third GEM are less concentrated compared to those of the fourth GEM, making it a positive ion blocker from the last stage of the gain element. At the same time, the energy resolution remains constant at around 11$\%$. 
In order to investigate these observations further, an additional study was performed by operating the quadruple GEM detector with the transfer and induction gap fields at the lowest IBF configuration obtained in Fig.~\ref{fig:ibf_vs_gapfld_dvgemsame}. The drift gap field was kept at 1 kV/cm. The gap fields used in this study are tabulated in Table~\ref{table:3}.\\ The results are shown in Fig.~\ref{fig:ibf_vs_gapfld_dvgemasymm_gapfldslowest}. There is a downward trend in IBF when each GEM layer is operated with higher potential across the GEM compared to the configuration in the previous GEM. That gives higher gain in each successive GEM. 
 
From Fig.~\ref{fig:ibf_vs_gapfld_dvgemsame} and Fig.~\ref{fig:ibf_vs_gapfld_dvgemasymm}, it can be seen that the effect of gap fields on IBF suppression is greater than operating each GEM with a different gain. By tuning to a specific gap field, IBF can be reduced to 6$\%$ as can be seen from Fig.~\ref{fig:ibf_vs_gapfld_dvgemsame}, whereas implementing only increasing GEM potential in successive GEMs can reduce the IBF from 42$\%$ to 34$\%$ as shown in Fig.~\ref{fig:ibf_vs_gapfld_dvgemasymm}.  
Tuning gap fields and the potentials on successive GEMs at the same time can reduce  IBF to a level of about 3$\%$ as is shown in Fig.~\ref{fig:ibf_vs_gapfld_dvgemasymm_gapfldslowest}.

The studies in Figs~\ref{fig:ibf_vs_gapfld_dvgemsame}-~\ref{fig:ibf_vs_gapfld_dvgemasymm_gapfldslowest} were performed in  ArCO$_{2}$(70:30) gas. Taking the optimized gap fields and GEM potentials from this study, the role of changing the drift gap field was further investigated for all gas mixtures listed in Table~\ref{table:2}. 
\begin{figure}[!htb]
\begin{center}
\includegraphics[width=0.70\textwidth] {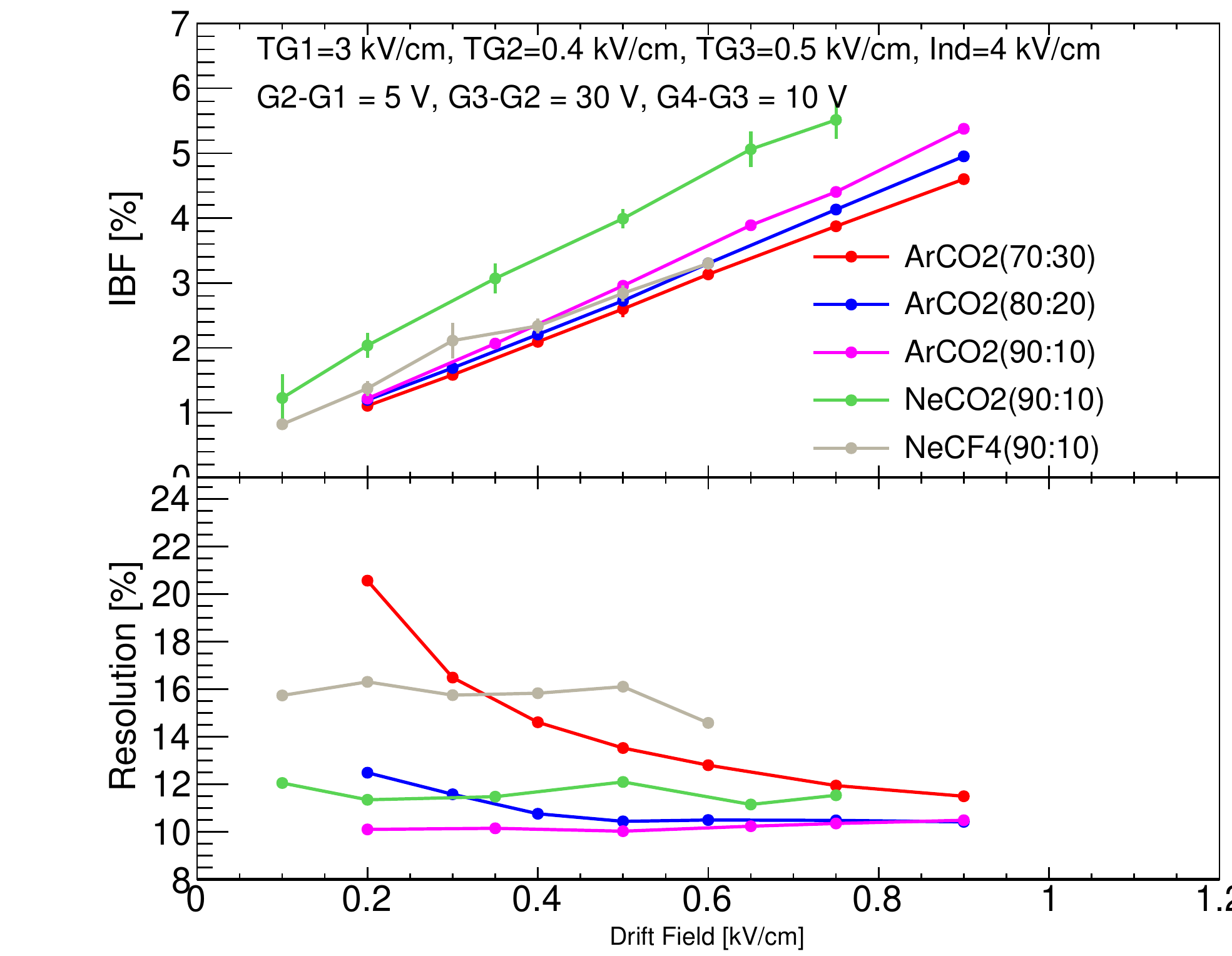}
\caption{ For the gas mixtures listed, the effect of the drift field on IBF and energy resolution of the quadruple GEM detector operated with gap fields and potentials on individual GEMs optimized for the lowest IBF is shown.} 
\label{fig:ibf_res_vs_drfld_lowestibdgapfld}
\end{center}
\end{figure}

Figure~\ref{fig:ibf_res_vs_drfld_lowestibdgapfld} shows the effect of the drift gap field on IBF and energy resolution when the quadruple GEM detector is optimized for lowest IBF by adjusting the transfer gap fields, induction gap field, and gain in each GEM. It can be seen that lowering the drift gap field can reduce the IBF down to 1\% for all studied gas mixtures. However, lowering the drift gap field worsens the energy resolution of the detector for the ArCO$_{2}$(70:30) gas mixture. For gas mixtures with a smaller amount of quenching gas, the drift field has no effect on the energy resolution. This may be attributed to the fact that with a higher concentration of quenching gas there is  lower electron transparency through the first GEM. This results in substantial deterioration of the energy resolution. Low electron transparency at low drift field with an increase in quenching gas has also been seen in other studies, for example in ~\cite{SAULI2006269}. The frequency of collisions between primary ionization electrons and quencher molecules is reduced as the quenching gas concentrations are reduced, and hence there is a negligible effect on the energy resolution of the detector. \\

Additional studies were performed to understand the effect of the operating gas mixture on IBF suppression when the detector is operated at a specified effective gain.  For this study Fig.~\ref{fig:an_cat_ibf_eps}(d) , which shows the correlation between $\epsilon$ and effective gain,  was fitted with a straight line. The fit parameters are tabulated in Table~\ref{table:4}. The slopes of $\epsilon$ versus effective gain are the same in   ArCO$_2$(90:10) and NeCO$_2$(90:10).

\begin{table}[h!]
\centering
 \begin{tabular}{||c | c |  c ||} 
 \hline
 Gas mixtures & Slope & Intercept  \\ [0.5ex] 
 \hline\hline
 ArCO$_2$(70:30) & 0.0061 $+/-$ 0.10 & 268.010 $+/-$ 27.88\\ 
 \hline
  ArCO$_2$(80:20) & 0.1632 $+/-$ 0.10 & 487.52 $+/-$ 66.18 \\
 \hline
 ArCO$_2$(90:10) & 0.2047 $+/-$ 0.01  & 357.68 $+/-$ 53.80 \\ 
 \hline
 NeCO$_2$(90:10) & 0.2054 $+/-$ 0.005 & 165.59 $+/-$ 28.72 \\ 
 \hline
 NeCF$_4$(90:10) & 0.1373 $+/-$ 0.005 & 173.85 $+/-$ 23.57 \\[1ex]
 \hline
 \end{tabular}
 \caption{Fit parameters of $\epsilon$ vs effective gain correlation from Fig.~\ref{fig:an_cat_ibf_eps}(d) for various gas mixtures used in this study. }
\label{table:4}
\end{table}

\begin{figure}[!htb]
\begin{center}
\includegraphics[width=0.70\textwidth] {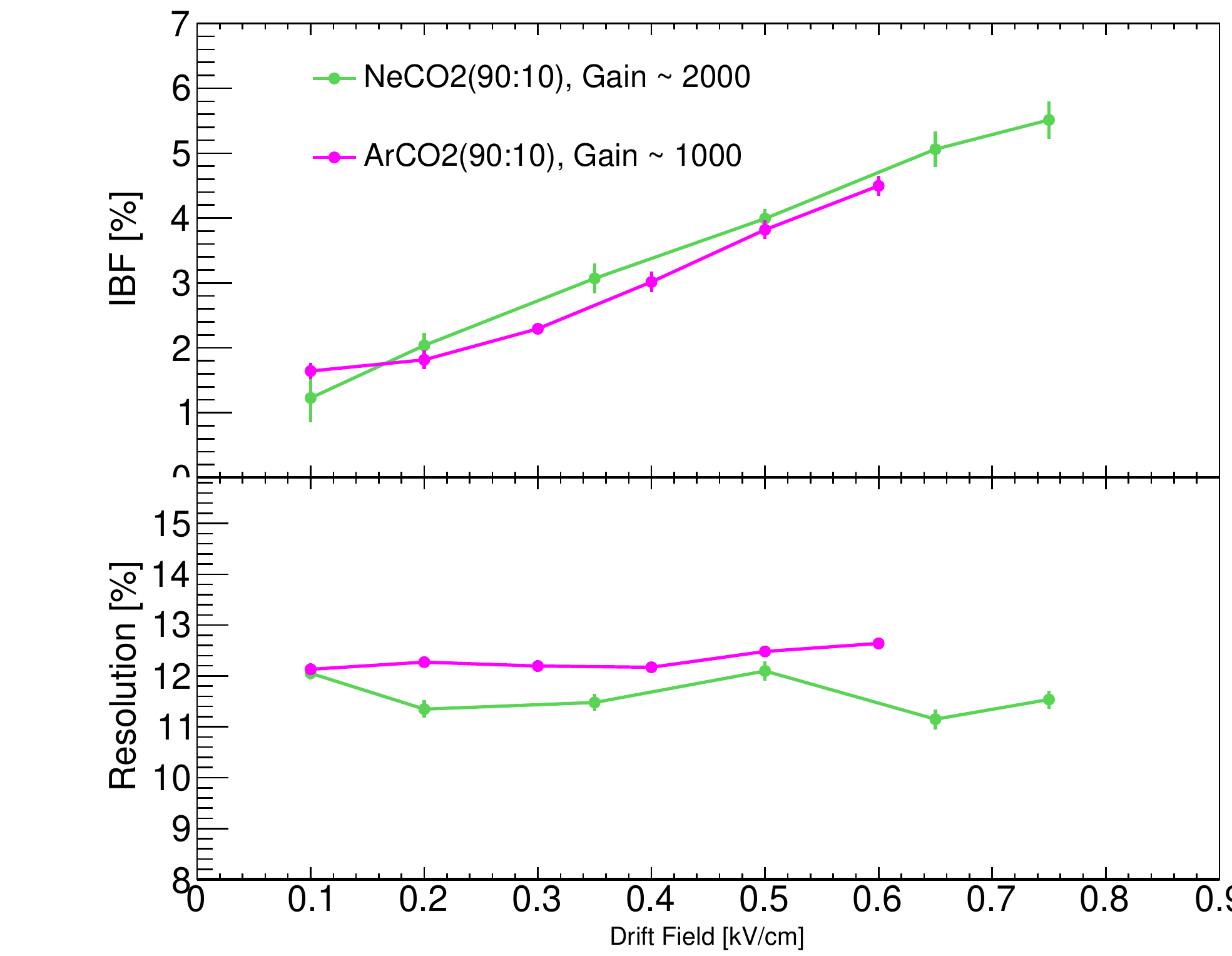}
\caption{ IBF and energy resolution vs drift gap field for quadruple GEM detector using ArCO$_{2}$(90:10) and NeCO$_{2}$(90:10) with the same initial $\epsilon$. The other gap fields and potential across each GEM were optimized for the lowest IBF configuration.  } 
\label{fig:ibf_res_vs_drfld_90arne_10co2}
\end{center}
\end{figure}

Both of these gas mixtures have the same CO$_{2}$ content. However for ArCO$_2$(90:10) the intercept fit parameter is twice that of NeCO$_2$(90:10). This shows that for the same effective gain, the number of back-flowing ions per incoming electron is twice as large in ArCO$_{2}$(90:10) as compared to NeCO$_{2}$(90:10). Thus, to understand the effect of ionizing gas on IBF suppression and energy resolution, the quadruple GEM detector was operated in NeCO$_{2}$(90:10) with an effective gain of twice the value used in  ArCO$_{2}$(90:10). Operating the detector under this condition will ensure that $\epsilon$ is same for these two gas mixtures. The effective gain of the detector in ArCO$_{2}$(90:10) was set to 1000 while for NeCO$_{2}$(90:10) it was set to 2000. Figure~\ref{fig:ibf_res_vs_drfld_90arne_10co2} shows IBF and resolution for the quadruple GEM detector in ArCO$_{2}$(90:10) and NeCO$_{2}$(90:10) at effective gains of 1000 and 2000, respectively, as a functions of the drift gap field while other operating parameters were tuned to the lowest IBF. Within statistical uncertainties, IBF can be suppressed at the same level for both of these gas mixtures while maintaining the same energy resolution provided the initial numbers of back-flowing ions per incoming primary ionization electron are the same. This shows that gas mixtures with the same $\epsilon$ do not affect IBF suppression; the operating voltages across different electrodes of the detector are the most important factor for suppressing the IBF.

\section{Conclusions}
This article presents a systematic study of ion back flow in quadruple GEM detectors in ArCO$_2$(70:30/80:20/90:10), NeCO$_2$(90:10) and NeCF$_4$(90:10) gas mixtures. It is shown that by optimizing the transfer gap and induction gap fields along with operating each of the GEMs at a different gain, one can reduce IBF from a few tens of a percent to about 6 percent at an effective gain of 2000. Further, by varying the drift field along with optimizing other gap fields and the potential across each GEM, the IBF can be reduced to the level of 1$\%$. 

The effect of the drift field on the energy resolution of the detector with optimized gap fields and potential across each GEM is also presented in this paper. A gas mixture having a greater fraction of quenching gas shows substantial degradation of the energy resolution with a decreasing value of the drift field, which is also the region of lowest IBF. Thus there is a trade-off between the lowest IBF configuration and the best energy resolution. \\

\section{Acknowledgements}
\label{sec:akn}
The authors are thankful to Bob Azmoun of Brookhaven National Laboratory and the late Dr. Richard Majka of Yale University for numerous discussions and suggestions on various aspects of the experimental set-up. The work presented in this article is supported in part by DOE grant number DE-FG05-92ER40712.



\bibliographystyle{elsarticle-num}
\bibliography{Quadgem_IBF_final}

\end{document}